\documentclass[showpacs,twocolumn,preprintnumbers,amsmath,amssymb,pre]{revtex4}

\usepackage{graphicx}
\usepackage{dcolumn}
\usepackage{bm}
\graphicspath{{./}{./figs/}}

\begin{document}

\title{Complex networks generated by the Penna bit-string model: Emergence of small-world and assortative mixing} \thanks{Phys. Rev. E 72, 045102, 2005.}
\author{Chunguang Li$^{1,2}$}
\email{cgli@uestc.edu.cn}
\author{Philip K. Maini$^2$}
\email{maini@maths.ox.ac.uk} \affiliation{$^1$Centre for Nonlinear
and Complex Systems, School of Electronic Engineering, University
of Electronic Science and Technology of China, Chengdu,
610054, P. R. China.\\
$^2$Centre for Mathematical Biology, Mathematical Institute,
University of Oxford, Oxford, OX1 3LB, United Kingdom.}
\begin{abstract}
The Penna bit-string model successfully encompasses many phenomena
of population evolution, including inheritance, mutation,
evolution and ageing. If we consider social interactions among
individuals in the Penna model, the population will form a complex
network. In this paper, we first modify the Verhulst factor to
control only the birth rate, and introduce activity-based
preferential reproduction of offspring in the Penna model. The
social interactions among individuals are generated by both
inheritance and activity-based preferential increase. Then we
study the properties of the complex network generated by the
modified Penna model. We find that the resulting complex network
has  a small-world effect and the assortative mixing property.
\end{abstract}
 \maketitle

In recent years, the usage of computational models has turned into
a major trend in the discussion of problems in population dynamics
and evolutionary theory. In 1995, a bit-string computer simulation
model for population evolution was introduced by Penna [1], which
encompasses the inheritance, mutation, evolution and ageing
phenomena. The Penna model has been so successful that it has
rapidly established itself as a major model for population
simulations.

On the other hand, complex networks composed of a large set of
interconnected vertices of various kinds are ubiquitous in nature
and society [2]. Examples include the Internet, the World Wide
Web, communication networks, food webs, biological neural
networks, electrical power grids, social and economic relations,
coauthorship and citation networks of scientists, cellular and
metabolic networks, etc. In recent years, many properties of
complex networks have been reported in the literature. Notably, it
is found that many complex networks show the small-world property
[3], which implies that a network has a high degree of clustering
as in a regular network and a small average distance between
vertices as in a random network. Another significant recent
discovery is the observation that many large-scale complex
networks are scale-free [4]. This means that the degree
distributions of these complex networks follow a power law form
$P(k)\sim k^{-\gamma}$ for large $k$, where $P(k)$ is the
probability that a vertex in the network is connected to $k$ other
vertices and $\gamma$ is a positive real number determined by the
given network. It is also found that many social networks exhibit
assortative mixing, the tendency for vertices in networks that
have many connections to be connected to other vertices with many
connections [5, 6].

If we consider social interactions among individuals in the Penna
model, the individuals (vertices) and the social interactions
(links) will form a complex network. In this paper, we first make
some modifications on the Penna model. We modify the Verhulst
factor to control the birth of individuals in the population, and
introduce activity-based preferential reproduction of offspring in
the Penna model, that is, the individuals with higher activity
will have higher probability to be selected to reproduce
offspring. The social interactions among individuals are generated
by both inheritance and activity-based preferential establishment,
that is the offspring will inherit a part (or all) of their
parent's social interactions at birth, and when the individual
becomes mature, if it has higher activity, it will have higher
probability to establish some new social interactions with other
individuals. Thus, with the evolution of the Penna model, it will
create a complex network with evolution and ageing, two common
properties observed in many real networks. Then we will study the
properties of the network generated by the modified Penna model
from the point of view of complex network theory.

The purpose of this paper is twofold. First, we try to mimic by
computer simulation the evolution of population and the formation
of social interactions in a society. Second, we provide a network
model with many important properties observed in real networks,
such as evolution, ageing, small-world effect and the assortative
mixing property.

We study the asexual Penna model in this paper. In the asexual
version of the standard Penna model, each individual is
represented by a bit-string of 32 bits (32 bits of 0 and 1), which
contains the information of when a hereditary disease will appear,
and plays the role of a chronological genome. Each bit corresponds
to a given age, and each individual can live at most for 32 time
steps (``years"). The presence of a 1 bit at a given position
means that the individual will suffer from the effects of a
genetic disease in that and the following years. The rules for the
individual to stay alive are: (i) the age of the individual is
less than or equal to 32; (ii) the number of inherited diseases
already taken into account at current age is lower than a
threshold $T$; and (iii) due to the restriction of space and food,
at each time step, the individual will stay alive with probability
$V=1-N(t)/N_{max}$ (the Verhulst factor), where $N_{max}$ is the
maximum population size allowed by the environment and $N(t)$ is
the current population size. There is a minimum reproduction age
$R$, after which the individual can generate offspring. The genome
of the offspring is a copy of the parent's genome, with some
random mutations. On short-time scales, nearly all mutations are
bad, so only deleterious mutation is considered: if a 0 bit is
chosen in the parent's genome, it is mutated to 1 in the offspring
genome; while if a 1 bit is chosen in the parent's genome, it
stays 1 in the offspring genome.

The Verhulst factor in the Penna model takes the idea from the
population growth model introduced by Verhulst in 1844. In that
model, an environmental carrying capacity $N_{max}$ is introduced
and the growth rate of a population is given by
\begin{equation}
\frac{dN}{dt}=rN(1-\frac{N}{N_{max}}),
\end{equation}
where $r$ is the intrinsic relative growth rate.

However, the Verhulst factor for controlling the population size
in the Penna model is too severe. Even if the population size is
much smaller than the environmental carrying capacity $N_{max}$,
this would predict that many individuals die in each year ``due to
the limit of space and food". The existing simulation results
indicate that the population size is usually very small compared
to $N_{max}$ (in [7], the simulation results show that the
population size never even reaches $0.3N_{max}$), and the
individuals usually die at a very young age (in [8], the
simulation results show that almost all the individuals die below
the age of 20). These results are mainly ``due to the restriction
of space and food'', although there is (more than) enough space
and food. It should be noted that some other researchers also
argued that the Verhulst factor is too severe in the Penna model
[9].

In fact, it seems from Eq. (1) that it is the population growth
rate rather than the rate of individuals kept alive in each time
step (in the Penna model, it is the latter case) that is
proportional to the Verhulst factor. Here we use a modified form
of Verhulst factor in the Penna model. We assume that an
individual will die only when its age reaches 32 or the number of
active diseases reaches the threshold $T$. We also assume that in
each year, the mature individuals will produce
$int\{N(1-N/N_{max})\}$ offspring. It is easy to show that the
population size will approach, but never exceed, $N_{max}$, if the
initial population size is $N_0\leq N_{max}$.

In the standard Penna model, all the mature individuals have the
same probability to reproduce offspring. But, in fact, the
reproductive activity is related to the individual's health
condition and age. Usually when exceeding the minimum reproductive
age, the young and healthy individual has higher reproductive
activity. In this paper, we use the following function to express
the dependence of activity on age
\begin{equation}
A^1=F(a)=\left\{\begin{array}{ll}0, & 0<a<R\\
1-\mbox{exp}[(a-32)/2],&R\leq a \leq 32\end{array}\right.,
\end{equation}
where $a$ is the age of an individual. We use the following
function to express the dependence of activity on health
\begin{equation}
A^2=H(m)=\mbox{exp}(-m/2), \hspace{1cm} 0\leq m<T,
\end{equation}
where $m$ is the number of active diseases at current age. We
assume that the total activity depends on the product of $A^1$ and
$A^2$, that is the higher the value of $A=F(a)H(m)$, the higher
the probability of the individual to be selected to reproduce
offspring. It should be noted that in [6, 10], the authors also
considered the health-controlled birth rate in the Penna model.

If we consider social interactions among individuals in the Penna
model, the individuals (vertices) and the social interactions
(links) will form a complex network. We summarize the evolution
rules for the networks studied in this paper as follows:

(i) Initialization: We start from $N_0\,(1\ll N_0\leq N_{max})$
individuals. Randomly, each individual has $[1,T]$ diseases at
randomly selected positions in the bit-string and each individual
is at the age in the range of $[1,32]$. The individuals are
randomly interconnected with probability $p_1$, where $p_1$
satisfies the inequality $p_1\gg \frac{\mbox{ln}(N)}{N-1}$ , so
that the resulting random network (random graph) is fully
connected [11]. In the following evolution, we assume that the
connection density is approximately fixed, that is the total links
in the network is approximately equal to
$int\{p_1\frac{N(N-1)}{2}\}$.

(ii) Death: In each time step, the individuals with age larger
than 32 (have reached the age 33) or with the number of active
diseases having reached the threshold $T$ will die, and the dead
individuals and all the interactions connected to them will be
removed.

(iii) Creating new interactions: In each time step, we create a
small number of new social interactions among existing
individuals. We assume the number of newly created interactions is
$int\{p_2[N(N-1)]\}$ with $p_2$ being a small probability. When
choosing two existing individuals to which a new interaction is
connected, we assume that two individuals are chosen from among
all existing ones, with the probability $\Pi(i,j)=\frac{A_i
A_j}{\sum_{m,n}A_m A_n}$, where $A_i$ is the activity of
individual $i$. That is, if two individuals both have high
activities, they will have a high probability of establishing a
new social interaction between them.

(iv) Birth: In each time step, the existing mature individuals
will reproduce $int\{N(1-N/N_{max})\}$ offspring. We choose an
individual to reproduce offspring with the probability
$\Pi(i)=\frac{A_i}{\sum_jA_j}$. That is, the higher the activity
of an individual, the higher is the probability to be selected to
reproduce offspring. This is the effect of activity-based
\emph{preferential reproduction}. The offspring genome is a copy
of the parent's one, and with probability $p_3$ a deleterious
mutation will occur.

An offspring will establish an interaction with the parent, and
each social interaction of the parent will be inherited with
probability 0.9 if the current total interactions $L>
int\{p_1\frac{N(N-1)}{2}\}$, and with probability 1 if $L\leq
int\{p_1\frac{N(N-1)}{2}\}$.

\emph{Remark 1}: In Rule (ii), in principle, there is a
possibility that some individuals become isolated due to the
removal of links. We assume that isolated individuals without any
social interactions cannot survive, so we simply remove these
individuals.

\emph{Remark 2}: Rule (iii) is motivated by the observation that
usually individuals with higher activity have more social
activities, and thus have more opportunities to establish new
social interactions with others. This is a kind of activity-based
\emph{preferential establishment of social interactions}. If there
is already a link between two selected individuals, then we will
do nothing.

\emph{Remark 3}: In our simulations, we found that if the
offspring inherit all the interactions of their parents, the
population will evolve to an almost fully connected network, but
if the offspring inherit each interaction with probability smaller
than 1, say 0.9, the number of interactions will decrease
gradually. So, we design an adaptive inherited rule in (iv) to
make the total number of interactions approximately equal to
$int\{p_1\frac{N(N-1)}{2}\}$.

The above rules are illustrated by numerical simulations as
follows. For a wide range of parameters, we can obtain similar
results, and we present some representative results here. We
consider a network with parameters $N_0=1000, N_{max}=1000, R=8,
T=4, p_1=0.1, p_2=0.0001, p_3=0.1$. We run the simulation several
times, and show representative results here. In principle, there
is a probability that the resulting network becomes unconnected,
but with these parameters this situation does not occur in our
simulations. The above rules are iterated 20,000 times to reach a
stationary population. In several independent runs, we found that
the resulting networks always have 998 or 999 vertices, which are
only slightly different from $N_{max}$. In the following, we
sometimes do not differentiate the resulting network size and
$N_{max}$ explicitly.

We next analyze some properties of the resulting network. We first
calculate the average path length (APL) [2] of the network. The
APL is obtained by averaging the APLs of networks generated by 10
independent runs with the above parametric values (in fact, the
APLs are nearly identical in each run). The value of APL for the
network is 2.0349, which is very small compared to the network
size. For the purpose of comparison, we also computed the APL for
networks with $N_{max}$ in the range $[100, 1000]$ (other
parameters are the same), and we found that the average path
lengths are all small and decrease as the network size increases
from 100 to 1000 vertices (Fig. 1 (a)).

The evolution of the clustering coefficient [2] of the network is
plotted in Fig. 1 (b). We can see from this figure that the
clustering coefficient increases rapidly from a small value (in
the initial random network, the clustering coefficient is
approximately equal to $p_1$) to a large value, and the clustering
coefficient fluctuates around about 0.45.
\begin{figure}[htb]
\centering
\includegraphics[width=4cm]{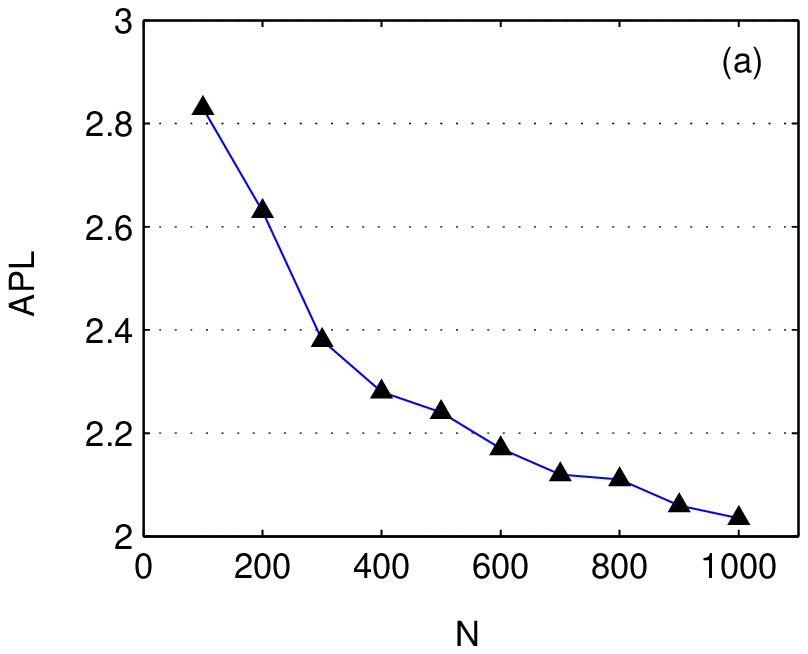}
\includegraphics[width=4cm]{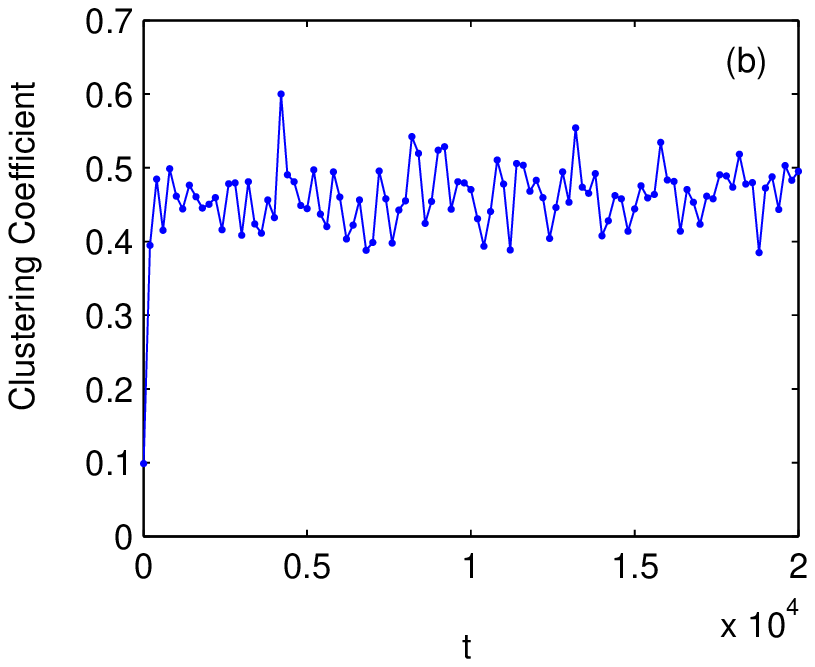}
\caption{(Color
  online) (a) The network average path length (APL) decreases as
network size increases from 100 to 1000 vertices. (b) The
evolution of the clustering coefficient.}
\end{figure}

As we know, ``small-world networks" are characterized by a high
degree of clustering and a small average path length which scales
logarithmically with the number of vertices [2]. We can see from
the above analysis that the network has a large clustering
coefficient and a small average path length, and the average path
length decreases with the increases of the network size, so the
network is a small-world network.

We plot the degree distribution [2] of the resulting network in
Fig. 2. The distribution looks like a Poisson distribution, but it
is ``fatter" on the right hand side of the peak than on the left
hand side. In the above rules, the reproduction of offspring can
be seen as a kind of preferential attachment in complex network
theory (the increase of new interactions is also a kind of
preference). But the offspring genome is mainly a copy of the
parent's one: if the parent has good health, the offspring will
also have good health (and high activity on becoming mature) with
a high probability, and this will increase the number of
individuals with high activity, so in the attachment the
preferential effect will be decreased and the random effect will
be increased. Due to the effect of mutation, the offspring who has
a healthy parent will lose good genome with a certain probability,
which will also increase the randomness of the attachment. So, the
degree distribution should be a result of the combination of
preferential attachment (power-law distribution) and random
attachment (Poisson distribution). This explains the degree
distribution in Fig. 2. It is known that the distribution of many
real networks can also be seen as a combination of power-law and
Poisson distributions.
\begin{figure}[htb]
\centering
\includegraphics[width=4cm]{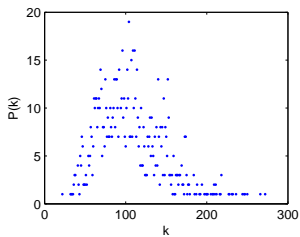}
\caption{(Color
  online) The degree distribution}
\end{figure}

Assortative mixing is also an important property of many networks,
especially social networks. A network is said to show assortative
mixing if the vertices in the network that have many connections
tend to be connected to other vertices with many connections [5,
12]. Assortative mixing can have profound effect on the properties
of a network. For example, it is found that assortative networks
are more robust to removal of their highest degree vertices than
the disassortative ones. To measure the assortative property of a
network, we can calculate the assortative coefficient. For mixing
by vertex degree in an undirected network, the assortative
coefficient is [5, 6]
\begin{equation}
r=\frac{M^{-1}\sum_i j_i k_i-
[M^{-1}\sum_i\frac{1}{2}(j_i+k_i)]^2}{M^{-1}\sum_i
\frac{1}{2}(j_i^2+k_i^2)-[M^{-1}\sum_i \frac{1}{2}(j_i+k_i)]^2},
\end{equation}
where $j_i, k_i$ are the degrees of the two vertices at the ends
of the $i$th edge, with $i=1,2, \cdots,M$. The coefficient is in
the range $-1\leq r\leq 1$, and if $r$ is positive, the network is
assortative.

For a network with $N_{max}=500$ created by the above evolution
rules [13], we calculate the value of $r$ according to Eq. (4),
and obtain that $r=0.4952$, which indicates strong assortative
mixing in this network. The scatter plot of the degrees of pairs
of vertices at the two ends of links is shown in Fig. 3 (a), and
the histogram of the degree differences between pairs of vertices
at the two ends of links is shown in Fig. 3 (b). From this figure,
we see that the network shows assortative mixing. In fact, in the
evolution of the population, there are individuals removed in each
time step, so the evolution of the network is robust to the
removal of vertices (including highest degree vertices), and has
self-repair ability.
\begin{figure}[htb]
\centering
\includegraphics[width=4cm]{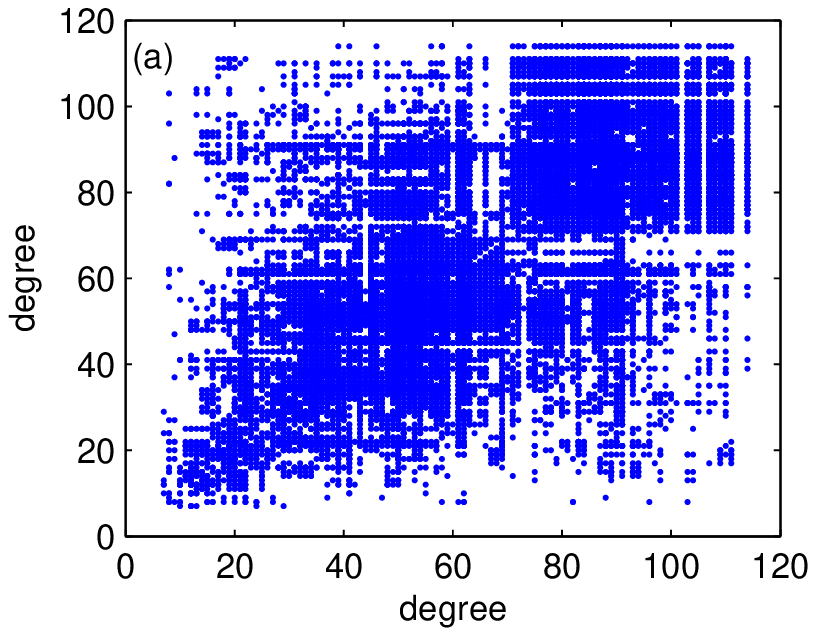}
\includegraphics[width=4cm]{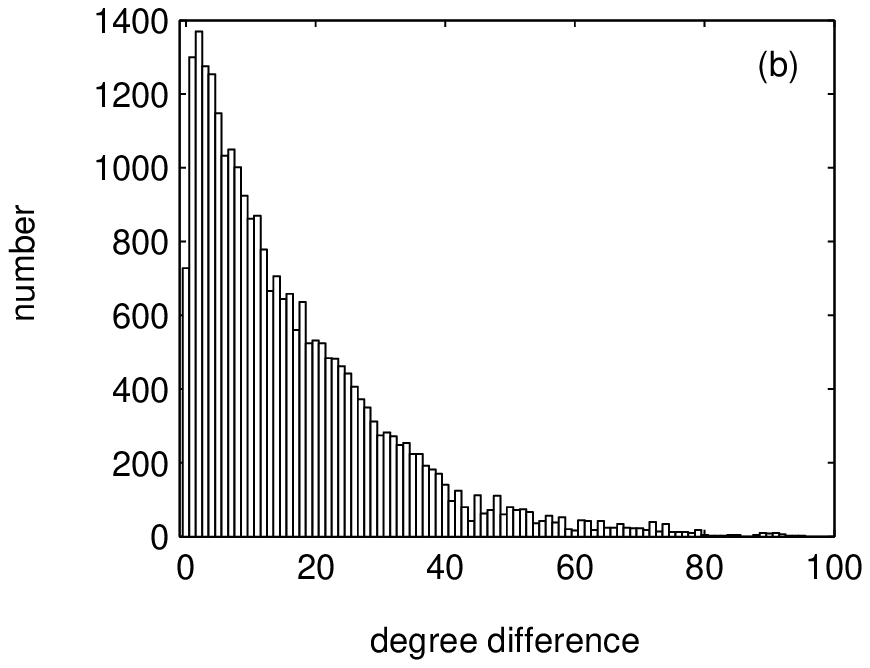}
\caption{(Color
  online) (a): Scatter plot of the degrees of pairs of vertices at
the two ends of links. (b): The histogram of the degree
differences between pairs of vertices at the two ends of links.}
\end{figure}

In summary, in this paper, we studied the evolution of a modified
Penna model, and showed that when taking the social interactions
into account, the individuals form a complex network. The
properties of the network were studied by computer simulation.
Numerical results indicate that the network has emergent
small-world and assortative mixing properties. The replication
rule is responsible for the small world effect, and the Penna
model is responsible for other properties, such as limited node
number, ageing effect, and the degree distribution, of the
network. This paper, to some extent, mimics the evolution and the
formation of social interactions in a society and also provides a
network model with many properties observed in real networks, such
as evolution, ageing, small-world effect and assortative mixing
property. We only studied the asexual case of the Penna model in
this paper; the sexual case, as well as other properties [14] and
theoretical analysis of the generated network, will be the subject
of future research.


\begin{thebibliography}{20}
\bibitem{"[12]"}
In [6], Newman extended the concept of assortative mixing in
networks to ``the tendency for vertices in networks to be
connected to other vertices that are like them in some way". In
this paper, we only consider mixing by vertex degree, that is, we
use, in a narrow sense, the concept of mixing [5].
\bibitem{"[13]"}
For $N_{max}=1000$, there are too many points in the scatter plot
of Fig. 3, so for the purpose of clarity, we use the results of
$N_{max}=500$ here.
\end{thebibliography}
\end{document}